\documentclass[authoryear,preprint,12pt]{elsarticle}
\usepackage{amsmath}
\usepackage{graphicx}
\usepackage{ulem}

\newcommand{\sopa}{SUPA, School of Physics and Astronomy, University of Edinburgh, Edinburgh EH9 3JZ, UK}
\newcommand{\ppls}{School of Philosophy, Psychology and Language Sciences, University of Edinburgh, EH8 9AD, UK}

\newcommand{\llus}{Literature and Languages, School of Arts and Humanities, University of Stirling, Stirling, FK9 4LA, UK}

\usepackage{color}

\journal{Cognition}

\begin{document}

\begin{frontmatter}

\title{Word learning under infinite uncertainty}

\author[1]{Richard A.\ Blythe} 
\ead{r.a.blythe@ed.ac.uk}
\author[2]{Andrew D.\ M.\ Smith}
\author[3]{Kenny Smith}

\address[1]{\sopa}
\address[2]{\llus}
\address[3]{\ppls}

\begin{abstract}
{Language learners must learn the meanings of many thousands of words, despite those words occurring in complex environments in which infinitely many meanings might be inferred by the learner as a word's true meaning. This problem of {\it infinite referential uncertainty} is often attributed to Willard Van Orman Quine. We provide a mathematical formalisation of an ideal cross-situational learner attempting to learn under infinite referential uncertainty, and identify conditions under which word learning is possible. As Quine's intuitions suggest, learning under infinite uncertainty is in fact possible, provided that learners have some means of ranking candidate word meanings in terms of their plausibility; furthermore, our analysis shows that this ranking could in fact be exceedingly weak, implying that constraints which allow learners to infer the plausibility of candidate word meanings could themselves be weak. This approach lifts the burden of explanation from `smart' word learning constraints in learners, and suggests a programme of research into weak, unreliable, probabilistic constraints on the inference of word meaning in real word learners.}
\end{abstract}

 
 \end{frontmatter}

\section{Word learning and indeterminacy of meaning}

Children are prolific word learners, learning around 60,000 words by age 18 \citep{Bloom00}. Their prodigious word-learning abilities are even more remarkable when we consider some of the challenges facing the word learner, including the need to segment words from connected speech \citep{saffran_96_statistical}, to generalise word forms across speakers \citep{henderson_05_two}, and to identify the syntactic properties of those words  \citep{mintz_02_category}. In this paper we focus on another aspect of the word learning problem: inferring word meaning. Children will typically encounter words in a complex environment. How do they know what these words mean? Every time a word is used, there may be many meanings which a learner could infer as the word's true meaning: the learner will face {\it referential uncertainty}. As discussed below, a widespread  observation in the literature is that there are potentially {\it infinitely} many candidate meanings which would be consistent with any given situation of usage. This idea, which we will refer to as {\it infinite referential uncertainty}, is commonly attributed to \citet{Quine60}, although on our reading (discussed below) we think Quine's central point was rather different.  

Regardless of its provenance,  the infinite referential uncertainty hypothesis has been crucial in the development of two approaches to word learning, which differ in emphasis but are entirely compatible in content. One position emphasises the importance of heuristics which guide word learning, serving to reduce referential uncertainty and allow the learner to make accurate inferences about word meaning. These heuristics might include: exploiting the attentional focus of a speaker \citep{tomasello_86_joint}; the assumption that words refer to whole objects \citep{macnamara_72_cognitive}; using knowledge of the meaning of other words to constrain hypotheses about the meaning of a new word, for example by assuming that words have mutually exclusive meanings \citep{markman_88_childrens}; using argument structure and syntactic
context to constrain the meaning of new words \citep{gillette_99_human}. Heuristic-driven accounts emphasise how such constraints enable learners to eliminate uncertainty about word meaning and form good hypotheses about word meaning on even a single exposure to a word. 

In the strongest accounts, these heuristics are hypothesised to eliminate all uncertainty. However, the possibility that word learners may be confronted with {\it some} residual referential uncertainty even after these heuristics have done their work has driven a recent burst of interest in a second approach to word learning, emphasising integration of information across multiple exposures as a means for learning in the face of referential uncertainty. {\it Cross-situational learning} comes in various flavours, from the classic formulation provided by e.g. \citet{Siskind96} to associationist treatments \citep{Yu07} to more minimal accounts \citep{Smith11,Medina11}. For instance, in its most powerful instantiation \citep[e.g.][]{Siskind96}, cross-situational learning involves tracking the set of meanings which has been consistently inferred on every exposure to some target word: the word's true meaning should be a member of this set, which can be winnowed down across a series of exposures until it includes only the true meaning. Cross-situational learning accounts typically assume the presence of heuristics which serve to reduce referential uncertainty to manageable levels: rather than replacing heuristics,  the contribution of this research is to explore the extent to which word learning is possible even given some residual (i.e. non-zero, but typically small) referential uncertainty. 

Focussing on the interaction between heuristic and cross-situational approaches, in previous work \citep{BSS10} we applied mathematical techniques to quantify what residual level of referential uncertainty a cross-situational learner can tolerate and still learn a large lexicon in a reasonable timeframe. Our previous work focussed on calculating learning times for lexicons given finite meaning spaces and finite levels of referential uncertainty. In this paper we apply similar techniques to tackle the problem of cross-situational learning for infinite meaning spaces under infinite referential uncertainty. In doing so, we seek to address what is often (perhaps rather loosely) called ``Quine's Problem" or ``the gavagai problem", the notion that word learning under infinite referential uncertainty is impossible. We show that word learning under such conditions is in principle possible, provided that learners have heuristics which at least rank the plausibility of each candidate meaning at every exposure. Thus, as in fact envisaged by Quine in his exposition of the indeterminacy of translation, word learning is possible if learners know, of the infinitely many possible meanings a word could have on any given situation of usage, how plausible each of those meanings are, and that some are more plausible than others. Within this very general set of conditions, given enough time, cross-situational learning can be used to eliminate uncertainty. Furthermore, cross-situational learning will in principle be possible even if the learner's heuristics only impose very weak constraints on the ranking in terms of plausibility.  This work therefore suggests similar conclusions to our previous work exploring finite referential uncertainty: word learning heuristics can in principle be far weaker than previously suggested and still allow word learning --- in fact, those heuristics can be so weak as to admit {\it infinitely many} possible meanings on any given exposure to a word, which renders single-exposure word learning impossible. Importantly, we therefore directly overturn the commonly-held assumption that word learning is impossible in the face of infinite referential uncertainty. Furthermore, the fact that word-learning heuristics which provide only weak constraints on possible word meanings can nonetheless allow word learning has potential implications for our understanding of the heuristics and cognitive biases underpinning word learning, and therefore on the empirical research attempting to uncover those biases. Firstly, this moves the explanatory burden from `smart' inference by learners to `dumb' crunching of cross-situational statistics, therefore requiring us to assume less of word learners in terms of their ability to accurately infer word meaning. Second, word learning heuristics do not need to allow learners to make good guesses on a single exposure to a word, which is a standard diagnostic in experimental research: weaker, unreliable, probabilistic heuristics can also play a key role, and therefore merit investigation.

\section{`Quine's Problem': learning under infinite uncertainty}
\label{sec:quine}

Words are used in complex environments, and each word could label any part of that complex environment. Worse, words can label objects and events which are not perceivable to speaker or hearer (e.g. events which are spatially or temporally distant from the time of speaking).  And this is only considering the obvious possibilities --- words might have `strange' meanings (e.g. featuring disjunctions of the meanings of `normal' words, meaning for instance ``a spark plug {\it or} an elephant", ``happiness {\it or} the number 17", etc). This idea, commonly attributed to Quine's work on radical translation (of which more below), appeals to the notion that on any situation there will be {\it infinitely many} possible meanings that a novel word could have: 

\begin{quote}
``Even if we restrict ourselves to middle-sized objects \ldots we are stuck with Quine's problem, which is that children who hear a word and know that it refers to a rabbit are still faced with an indefinite number of possible meanings for this word" \citep[p. 56]{Bloom00}

``Quine (1960) points out that there are an infinite number of true facts about the world that a learner might need to entertain as potential meanings of each utterance." \citep[p. 45]{Siskind96}

``Worse, or so philosophers tell us, learners might conjure up absurd and endlessly differing representations for those entities we adults call `the cats.'" \citep[p. 136]{gillette_99_human}

``Famously articulated by Quine (1960), in any naming situation there are infinite interpretations for an unknown word. Thus, children face a daunting task of ambiguity resolution that they must solve thousands of times.'' \citep[p. 831]{mcmurray_12_word}

``Word learning is often described as a difficult task because the world offers infants a seemingly infinite number of word-to-world mappings in just one moment in time (Quine, 1960)." \citep[p. 375]{vlach_13_memory}

``Determining the meaning of a newly encountered word should be extremely hard, due to the (in principle, unlimited) referential uncertainty inherent in the task (Quine, 1960)." \citep[p. 480]{Smith11}
\end{quote}

Such claims about infinite referential uncertainty are widespread in the literature, and have played an important role in the development of the theoretical motivation for research on the heuristics children use to eliminate uncertainty during word learning: seminal papers on the Mutual Exclusivity constraint \citep{markman_88_childrens}, the shape bias \citep{landau_88_importance}, joint attention \citep{baldwin_91_infants}, or lexical constraints in general \citep{golinkoff_94_young}  make explicit reference to the problem of there being ``infinitely many" or ``limitless'' meanings a word could have. The consensus is that word learning is impossible given this infinite uncertainty, and that heuristics are required to eliminate some of these candidate meanings. In this paper we explore the validity of this widely-held and entirely reasonable intuition, and in particular show that it does not hold in a wide range of well-defined circumstances. However, before doing so it is worth briefly considering whether Quine's Problem was actually posed by Quine.

\section{Quine on word learning}
\label{sec:what_quine_said}

\citet{Quine60} introduces the problem not in terms of word learning, but in terms of  ``radical translation", his examination of how the language of a ``hitherto untouched people" can be translated. Much of his discussion focusses on issues which are only tangentially relevant to word learning (such as how we can understand signs for agreement or disagreement from our informant) or which are relevant to language learning in general but not to referential uncertainty (e.g. the order in which types of meaning must be learnt). The key passage, referred to implicitly in much of the literature, is as follows:

\begin{quote}
``For, consider ÔgavagaiÕ. Who knows but what the objects to which this term applies are not rabbits after all, but mere stages, or brief temporal segments, of rabbits? In either event the stimulus situations that prompt assent to ÔGavagaiÕ would be the same as for ÔRabbitÕ. Or perhaps the objects to which ÔgavagaiÕ applies are all and sundry undetached parts of rabbits; again the stimulus meaning would register no difference." \citep[p. 51-52]{Quine60}
\end{quote}

Interestingly, Quine never explicitly states that there are an infinite number of possible meanings for each word at any episode, although  this is an entirely reasonable inference from his discussion of what we referred to earlier as `strange' meanings. Rather, it seems to us that Quine is much more concerned with the fact that some candidate meanings are {\it in principle always indistinguishable} from a word's true meaning because they are always applicable to the same stimuli: ``Point to a rabbit and you have pointed to a stage of a rabbit, to an integral part of a rabbit, to the rabbit fusion, and to where rabbithood is manifested" \citep[p. 52]{Quine60}. As we note below, this renders word learning impossible in principle, if we treat word learning as a process of eliminating spurious meanings, as theories of cross-situational learning typically do; in other words, we accept Quine's central point that, if two candidate word meanings are always indistinguishable, cross-situational learning cannot distinguish between them. To escape this conundrum, we have to assume that some of these indistinguishable meanings (i.e. ``rabbit") are just more plausible than others (``undetached rabbit parts"). Quine suggests that we can't assume this {\it a priori}, although he acknowledges that in practice we do rank candidate meanings in terms of plausibility, saying that the linguist assumes that  ``the native is enough like us to have a brief general term for rabbits ... [and not] for rabbit stages or parts" (p. 52). In the remainder of this paper, we show that Quine's intuitions on plausibility ranking over candidate word meanings solves the problem of word learning in the face of infinite referential uncertainty, and quantify how weak those rankings can be.

\section{Model of the learning problem}

Following the approach taken by other modelling and experimental work on cross-situational learning, we characterise the word learning process as one of eliminating uncertainty about word meaning, and we seek to identify conditions under which the meaning of a word can be uniquely determined after multiple exposures.  We take an ideal observer approach \citep{frank_11_three} and consider an optimal cross-situational word learner attempting to identify the meaning of a single word; we measure time $t$ in terms of the number of exposures to this word that the ideal learner has received.  Each time the word is uttered, the learner infers (by applying their battery of word learning heuristics to the situation in which the word is used) one or more possible discrete meanings for that word drawn from an infinite set of meanings according to some specified probability distribution.  We make two assumptions regarding the form of this distribution:

\begin{description}
\item[Assumption 1:] The word's true meaning, the {\it target meaning}, is always included as one of the inferred meanings.
\item[Assumption 2:] All other meanings (the {\it incidental meanings}) have some nonzero probability of {\it not} being inferred.
\end{description}

Assumption 1 is often identified as a problem in theories of cross-situational learning \citep[see e.g.][]{Gleitman90} --- although it simplifies our analysis, we show in Section \ref{sec:target_failure} that word learning is possible if we violate Assumption 1 and allow that the word's true meaning is not always (or even not often) inferred. 

Assumption 2 corresponds to the requirement that not all possible incidental meanings are inferred on every exposure to a word, which would clearly render cross-situational learning impossible. Furthermore, it corresponds to (our reading of) Quine's concern, discussed above, that some meanings are in principle indistinguishable from the word's true meaning: were this the case, Assumption 2 would be violated for those meanings. We consider this assumption to be necessary for cross-situational word learning to be possible, and probably necessary for any theory of word learning which conceives of learning as a process of converging on a correct meaning at the expense of all alternatives.

These assumptions about the candidate meanings a learner infers on a single exposure to a word therefore constitute our model of the learner's word learning heuristics: manipulating the probability distribution over inferred meanings allows us to simulate learners with very powerful word learning heuristics (e.g. when the learner always infers the target meaning and a very small number of incidental meanings) or far weaker heuristics (e.g. where the number of incidental meanings inferred on any episode is large or even infinite, due to the learner's inability to make strong inferences about the word's true meaning). Note that, unless we assume that the learner's heuristics are maximally strong and eliminate {\it all} referential uncertainty (i.e. no incidental meanings are inferred, an extreme version of our Assumption 2), cross-situational learning still has to do some work: every exposure will feature multiple candidate meanings, and the target meaning will therefore not be identifiable on any single exposure.

Our model is agnostic about the source of these constraints on possible word meanings, but we envisage them as arising through the interplay between the learner's word learning heuristics and the (linguistic, social and physical) context in which a word is used: for instance, on hearing a word, a learner may assume that it is likely to be a noun given the syntactic context in which it occurred \citep{gillette_99_human}, that it is likely to refer to whole objects rather than part of an object \citep{macnamara_72_cognitive}, that it is likely to refer to one of the cluster of objects that the speaker was attending to \citep{tomasello_86_joint}, and that, given all the foregoing, it is likely to refer to the toy that just beeped and flashed in a rather salient manner. 

Other than these two assumptions given above, we do not place any restrictions on the meanings that are inferred or the relationships between the inferences that are made within or between exposures.  It should be noted that assuming discrete meanings, as we do here, does not commit us to any assumptions about the nature of meanings or the relationships between meanings. For instance, meanings could be interpreted as existing in a hierarchically- and similarity-structured space, which would be captured in our model by manipulating the set of incidental meanings associated with the word's true meaning and the distribution from which those incidental meanings are drawn, such that incidental meanings which are related to the target are more likely to be inferred by the learner, and more general meanings are more likely to be inferred than more specific, restricted meanings. We return to word learning for such structured meaning spaces in the discussion. Relatedly, in Sections \ref{sec:correlations_within_episodes} and \ref{sec:correlations_between_episodes} we explore scenarios where there are correlations between meanings, either within a single exposure (the fact that one meaning has been being inferred on a given situation could increase, or decrease, the probability that other meanings are inferred, for example if those meanings are related) or across exposures (the inference of a meaning in one exposure could increase or decrease the probability of its being inferred in subsequent exposures, for example as a result of the temporal structure of dialogue).

We assume that our idealised learner uses the full cross-situational learning strategy defined in e.g. \citet{Siskind96} or \citet{BSS10}, whereby the learner keeps track of the set of meanings that have been inferred in {\it every} exposure up to time $t$. Due to Assumption 1, the target meaning is always part of this set; if the sequence of exposures is such that this set comprises precisely one meaning (which must be the target meaning), the word is taken to be learnt. In Section \ref{sec:target_failure}, where we relax Assumption 1, we necessarily adopt a less restrictive, and consequently less powerful, conception of cross-situational learning.

\begin{figure}
\begin{center}
\includegraphics[width=\linewidth]{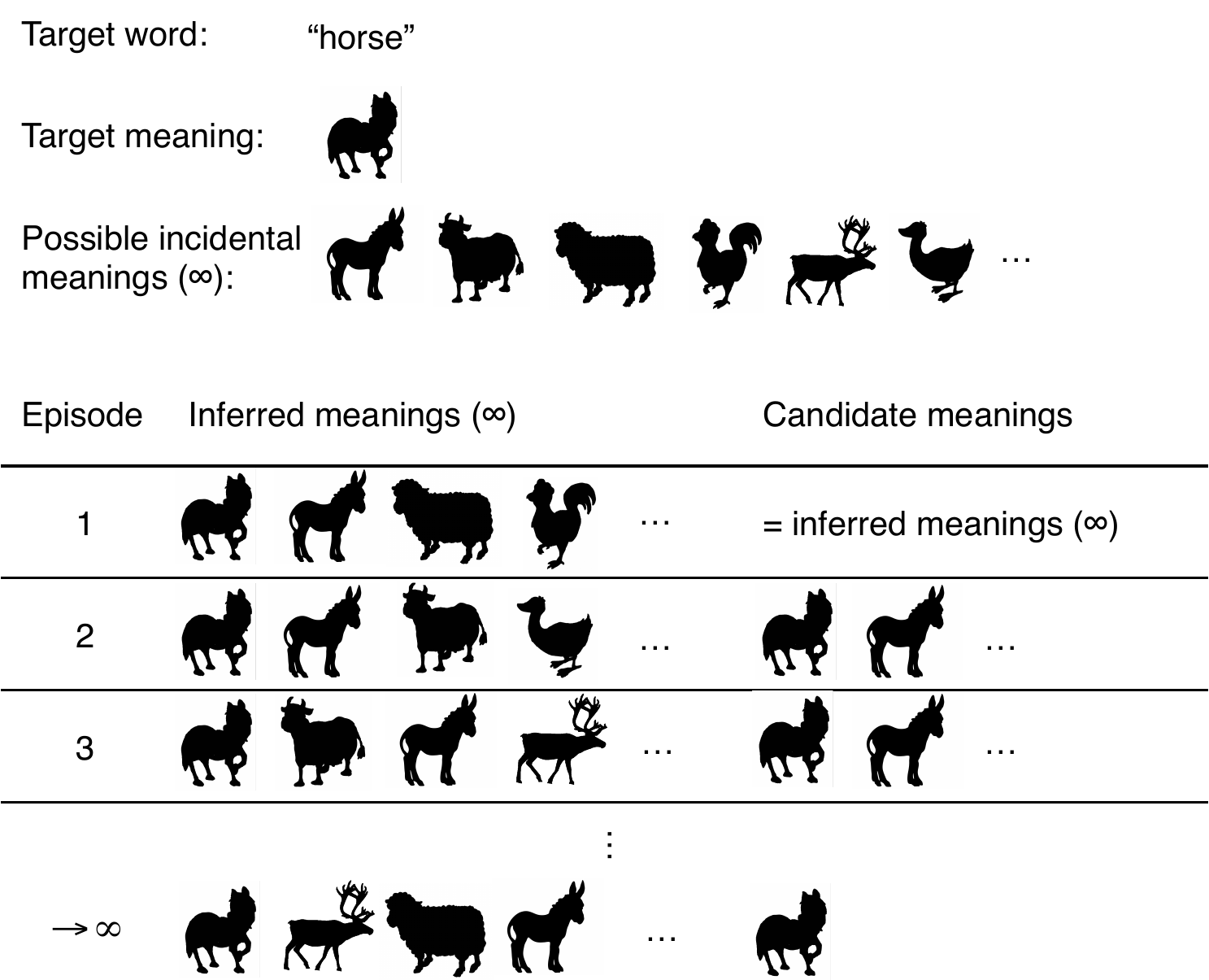}
\end{center}
\caption{\label{fig:learning_model} A sketch of our model of word learning. For every word (here, {\it horse}) there is a target meaning (represented here by the silhouette of a horse) and an infinitely-large set of possible incidental meanings (represented here by silhouettes of other animals). Learning consists of a series of episodes; on each episode  the learner samples meanings from the set consisting of the target meaning plus the incidental meaning, with these meanings being sampled with differing probabilities (e.g. if our Assumptions 1 and 2 are followed, the target meaning is always sampled, and every incidental meaning has some probability of {\it not} being sampled at any given episode). This sampling procedure represents the process whereby the learner infers likely word meanings based on their word learning heuristics and the context in which the target word is used. On the basis of each such learning episode, the learner updates their set of candidate meanings for the target word: on the first exposure this will simply be the set of inferred meanings, each of which is equally likely to be the word's true meaning; over episodes this set of candidate meanings is winnowed down as candidates fail to occur in the set of inferred meanings and can therefore be eliminated. Eventually, if cross-situational learning is possible, the set of candidate meanings will consist of a single meaning, the target meaning, and the word has been learnt.}
\end{figure}

Our model of word learning is sketched in Fig.~\ref{fig:learning_model}. Given this very simple model, what are the formal conditions under which word learning is possible in finite time? Given the standard interpretation of Quine discussed above, we are particularly interested in scenarios where words can be learned in finite time despite potentially infinite referential uncertainty.  We place no restrictions on how small learning times must be: in previous work \citep{BSS10} we investigated learning times for large lexicons in timescales which were comparable to those likely to be available for real language learners, but since here we are concerned with exploring the feasibility of cross-situational learning {\it in principle}, learning in finite time is our only requirement: learning in practicable timescales, as we explore for the finite meaning case in \citet{BSS10}, is an additional constraint and a worthwhile avenue for future work. 

\section{Formal conditions for word learning in finite time}

\subsection{A simple formulation of cross-situational learning}

We define the learning time $t^\ast(\epsilon)$ that is needed for the meaning of a word to be uniquely identified with probability $1-\epsilon$, where $\epsilon$ is some small constant.  A probabilistic notion of learning time is necessary because it is always possible to construct an arbitrarily long sequence of exposures where the target word remains unlearned (e.g. any sequence where the target meaning and one particular incidental meaning are both inferred: such sequences constitute temporary instances of the problematic case that concerned Quine, namely where two candidate word meanings cannot be differentiated).  The learning time is finite when the probability of such vexatious sequences is sufficiently small that $t^\ast(\epsilon)<\infty$ for any arbitrarily small, but nonzero, $\epsilon$. 

To understand this criterion in more detail, it is helpful to think of a population of learners, each drawing their own set of inferences of a word's meaning from a common distribution. Then, $\epsilon$ can be interpreted as the fraction of all learners who would not have learnt the target word's meaning at time $t=t^\ast(\epsilon)$. Clearly, if $t^\ast(\epsilon_0)$ becomes infinite at some finite $\epsilon_0$, then some fraction of learners fail to learn the word in a finite amount of time. Furthermore, once the population size $N$ exceeds $1/\epsilon_0$, then there is typically at least one agent in the population who fails to learn the word. Our criterion thus amounts to insisting that {\it all} agents learn the word in a finite amount of time (although this can be very long), and we consider it preferable to using other statistics, such as the mean time to learn the word (which can be infinite, even though every member of the population can learn the word in a finite amount of time) or the median (which can be finite even when nearly half the population fail to learn the word in a finite amount of time).

As outlined above, we assume that meanings are discrete, and that there are infinitely many possible incidental meanings (i.e. there are infinitely many meanings which {\it could}, in principle, be inferred on any occurrence of the target word). We therefore have a one-to-one mapping between positive integers $m$ and incidental meanings.  We define ${\cal K}(t)$ as the set of incidental meanings that have been inferred in {\it every} episode up to time $t$; the word is learnt when ${\cal K}(t^\ast)=\emptyset$, i.e. when all incidental meanings have been eliminated, and the word's true meaning is therefore revealed.  The crucial quantity in determining whether the learning time is finite or not is the {\it residual plausibility} of an incidental meaning $m$, which is defined as follows
\begin{equation}
\label{pmt}
 p_m(t) = \Pr\left[ m \in {\cal K}(t) | {\cal F}_{m-1} \cap  {\cal K}(t) = \emptyset \right]
\end{equation}
where ${\cal F}_m$ is the set of meanings $\{1, 2, \ldots, m\}$. In words: the residual plausibility of incidental meaning $m$ at time $t$ is the probability that it is still a candidate meaning for the word {\it given that} the first $m-1$ incidental meanings have all been excluded after $t$ exposures.  This definition can be applied for any ordering of the meanings; however it is often convenient to assume that they are arranged in decreasing order of their probability of being inferred in any given episode.

Given this definition, we can now state our main results, a formal derivation of which is given in the Appendix.  First, the probability $L(t)$ that the word is learnt by time $t$ is simply the probability that all incidental meanings have been eliminated, i.e. ${\cal K}=\emptyset$. We find that this can be written in terms of the residual plausibilities as
\begin{equation}
\label{Loft}
L(t) = \prod_{m=1}^{\infty} \left[ 1 - p_m(t) \right] 
\end{equation}
i.e. the probability of learning a word is given by the probability that the first non-target meaning has been excluded, multiplied by the probability that the second non-target meaning has been excluded conditioned on the first being excluded, multiplied by the probability that the third non-target meaning has been excluded conditioned on the first two being excluded, and so on down the infinite hierarchy of non-target meanings. 

Then, we find that the learning time $t(\epsilon)$ is finite if and only if 
\begin{equation}
\label{iff}
\lim_{t\to\infty} \left[ \sum_{m=1}^{\infty} p_m(t) \right] = 0 
\end{equation}
i.e. learning time is only finite if the residual plausibility of {\it all} incidental meanings reaches 0.

Informally, we can interpret this result as follows: if the combined residual plausibility of all incidental meanings vanishes in the limit of an infinite sequence of exposures, the meaning of a word can be identified with arbitrarily high probability in a finite time.  \citet{BSS10} considered cases where the set of possible incidental meanings was finite: as is made clear in the formulation above, the learning time in such cases will be finite as long as there is some probability that each incidental meaning is excluded at any given exposure, which follows from our Assumption 2. Therefore, under our assumptions, word learning is always possible when the set of possible incidental meanings is finite. The case of an infinite set of incidental meanings is best elucidated by means of explicit examples, to which we now turn.

\subsection{Learning times given infinite meaning spaces}

\label{sec:indinf}

Let us start by assuming that each incidental meaning $m$ is inferred with a constant probability $a_m$ in each exposure, and these inferences are statistically independent (that is, the probability that both meanings $m$ and $m'$ are inferred in any given exposure is just $a_m a_{m'}$, and is the same in each episode).  Under such circumstances, the probability that meaning $m$ is inferred in all of the first $t$ episodes is $a_m^t$, and this does not depend on whether any other meanings (e.g. meanings $1,2,\ldots,m-1$) have been inferred in each of those $t$ exposures or not.  In this case, the residual plausibility of meaning $m$ at time $t$ is  $p_m(t) = a_m^t$.  Now, since by our Assumption 2 all $a_m<1$, i.e. all incidental meanings have some non-zero probability of not being inferred on any given exposure, it follows that $p_m(t) \to 0$ as $t\to\infty$: that is, every incidental meaning becomes ever more implausible as time goes on.  However, unlike in the case of finite sets of incidental meanings, this does not in itself guarantee a finite learning time, as we shall now see.

Let us arrange the meanings in decreasing order of their inference probability, i.e., such that $a_{m+1} \le a_m$.  Suppose that, for all incidental meanings, their probability of being inferred on any given exposure follow a power law: $a_m = c m^{-\gamma}$, where $c$ and $\gamma$ are positive constants. $c$ denotes the probability of inferring the most frequent incidental meaning on any given episode, and must be less than 1 by our Assumption 2. $\gamma$ denotes the rate of decay of the power-law distribution: for high $\gamma$, $a_{m}$ decreases rapidly with $m$ (i.e. only the few most frequent meanings have any substantial probability of being inferred on any given trial); as $\gamma$ is reduced, the distribution of $a_{m}$ flattens out,  such that even lower-ranked meanings have a substantial probability of being inferred in any episode.  Under these conditions, the sum in (\ref{iff}) converges (i.e. approaches some constant) for any $t>1/\gamma$ and furthermore vanishes (i.e. converges to 0) in the limit $t\to\infty$.   Importantly, note that the rate of decay of the inference probabilities can be very slow, for example, if $\gamma$ is very close to zero.   Indeed, when $\gamma\le1$, the mean number of meanings inferred in each episode is {\it infinite}: the infinitely-long tail of meanings, each with substantial probability of being inferred at each instance of use of the target word, leads to a situation where the learner is confronted with infinite referential uncertainty;  nevertheless, the power-law decay of $a_m$ is still sufficiently fast for a cross-situational learner to be able to eliminate all non-target meanings in finite time. This refutes the common intuition, discussed above, that cross-situational learning cannot work in principle in the face of infinite referential uncertainty. 

\begin{figure}
\begin{center}
\includegraphics[width=\linewidth]{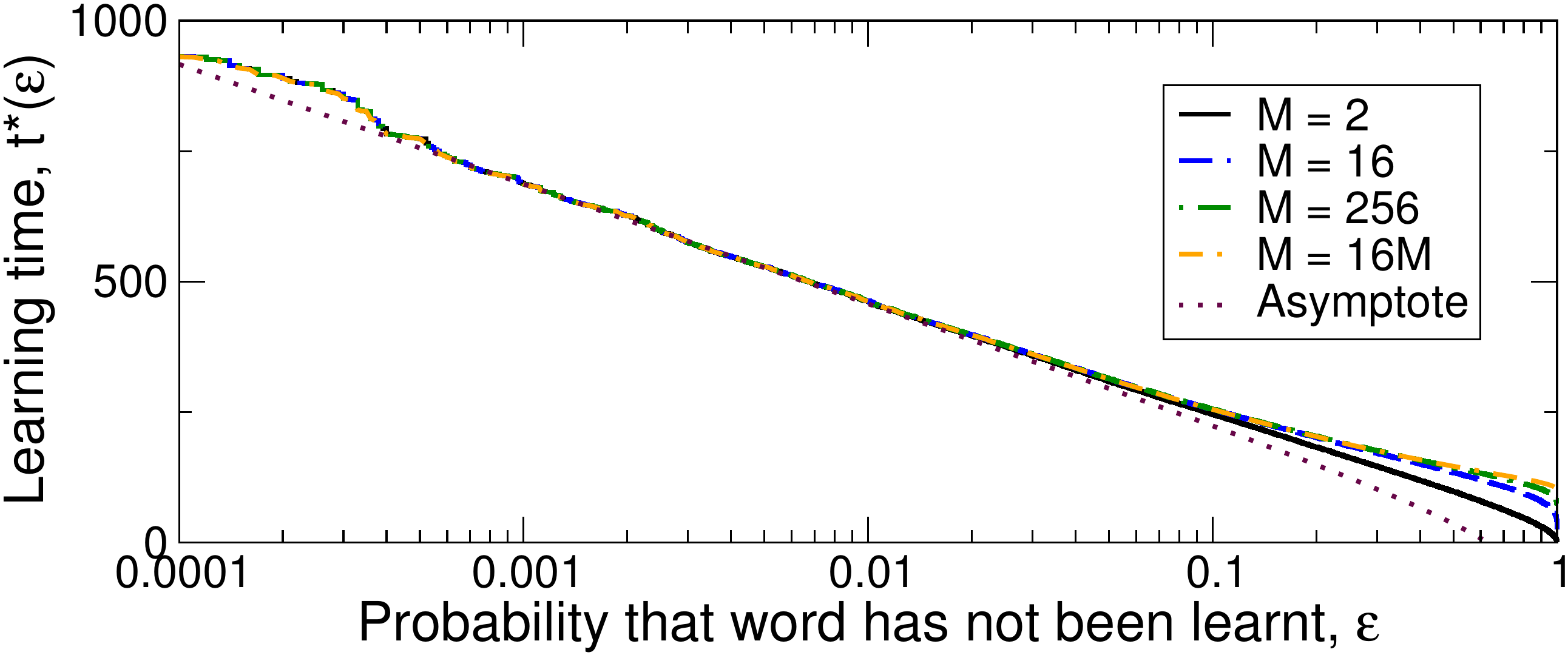}
\end{center}
\caption{\label{fig:oneword-power} Time to learn a word with probability $1-\epsilon$ with confounder frequencies governed by the power-law $a_m = c m^{-\gamma}$, with $c=0.99$ and $\gamma=0.01$.  Note that, crucially, learning times are independent of the number of incidental meanings $M$, as indicated by the invariance of learning time as we vary $M$. For $\epsilon<0.01$, the learning times are well approximated by the asymptotic result $t^\ast(\epsilon) \sim \ln(\epsilon)/\ln(c)$. The values of $M$ used in the simulation are powers of $2$: $2^1=2, 2^4=16, 2^8=256$ and $2^{24}=16,777,216$ (denoted $16$M in the figure).}
\end{figure}

This can be illustrated using simulations of cross-situational learning. Since it is impossible to use an infinite meaning space in a simulation, our approach here is to vary the number of available meanings, $M$, and investigate what happens as $M$ increases (using $c=0.99$ and $\gamma=0.01$). In Figure~\ref{fig:oneword-power}, we plot the time at which all but a fraction $\epsilon$ of learners have learnt the word: this time is independent of $M$ for sufficiently small $\epsilon$.

Note that for the exponent $\gamma=0.01$ used in the plot, the mean number of meanings inferred on any given episode is $M^{1-\gamma}$, i.e., $M^{0.99}$, meaning that virtually all meanings that could be inferred are inferred on every instance. Despite this, if we are only interested in the most laggardly learners, only the most frequent of these many incidental meanings is relevant to learning time: by the time the most problematic incidental meaning has been eliminated by the slowest learner, the other less frequent incidental meanings will also have been eliminated. The influence of the most frequent confounder on learning times is a feature of all the cases we consider in this paper. In fact, in the regime $\epsilon<0.01$, we find good agreement between our simulation results and a very crude approximation to (\ref{Loft}) in which we consider only the most frequent confounder, which is inferred on every episode with probability $c$, i.e.,
\begin{equation}
L(t) \approx 1 - a_1^t = 1 - c^{t} \;. 
\end{equation}
Setting this equal to $1-\epsilon$ we obtain an estimate for the learning time $t^\ast(\epsilon)$.  Here, we find that that, as $\epsilon\to0$, $t^\ast \sim \ln(\epsilon)/\ln(c)$.  As can be seen from Figure~\ref{fig:oneword-power}, the agreement with the numerical data for $\epsilon < 0.01$ is excellent. 

Given that we find that a word can be learnt even when the plausibility ranking of non-target meanings is very weak, one may legitimately ask whether it is sufficient only for $a_m$ that decreases to zero as $m\to\infty$ to guarantee a finite learning time. It turns out that this is not the case. In the Appendix we provide the example of a set of frequencies that decay logarithmically, and find that the word is not necessarily learnt in finite time. In this case, whether learning times remain small or diverge towards infinity depends on the size of the set of incidental meanings: for logarithmic decay, it turns out that learning times are determined by the most frequent confounder, as above, when this set is less than $10^{18}$, a vast set of possible incidental meanings.  Only above this size may learning times that diverge as $M$ is increased become apparent. This example shows that although a cross-situational learner can in principle fail to learn the meaning of a word in a finite time when the inference probabilities decay logarithmically, learning is nevertheless possible unless the number of meanings that may be inferred is extremely large.

\subsection{Correlated meaning inferences within episodes}
\label{sec:correlations_within_episodes}

It is instructive to consider cases where the inference of certain meanings enhances (or diminishes) the probability that certain other meanings are inferred: in our simple model of meaning, such correlations between meanings can be used to capture hierarchical- or similarity-based structure within the meaning space; as such, cross-situational learning in the presence of such correlations is an important topic of study. An extreme (but illustrative) case is where the inference of incidental meaning $m$ on any given episode implies that all meanings that are more likely to be inferred are also inferred.  If we order the meanings according to decreasing probability of being inferred, this means that whenever meaning $m$ is inferred, all meanings $m'<m$ are also inferred.  Conversely, if meaning $m$ is not inferred, all meanings $m'>m$ are also not inferred (see next paragraph for the relationship between such meanings and Quine's `strange' meanings).

Under this scenario, the residual plausibility of all meanings $m>1$ (i.e. other than the most probable incidental meaning) is zero.  This is because if all meanings $m'<m$ have been excluded, then meaning $m$ must also have been excluded.  Hence, the probability the word is learnt after time $t$ is just equal to the probability that the most probable incidental meaning has been excluded.  All that is needed for a finite learning time is for this probability to approach unity as $t\to\infty$.  These within-episode correlations therefore lead to {\it faster} learning than when meanings within a single episode are statistically independent. More generally, any incidental meaning $m$ whose inference implies the inference of a (necessarily more plausible) meaning has residual plausibility $p_m(t)=0$, and can therefore be ignored for the purposes of computing learning times.  All such {\it shadowed} meanings are therefore invisible to a cross-situational learner. Quine's `strange' meanings seem excellent candidates for shadowed meanings: if ``undetached rabbit parts" is inferred, then ``rabbit" would necessarily also be inferred and would be more plausible.

What about weaker within-episode correlations, which do not produce shadowed meanings? Learning times here should be somewhat similar to the case where each incidental meaning is statistically independent of any other, since weakening of within-episode correlations gradually approaches this statistically independent case.  While we do not have a general formation for these weak-correlation scenarios,  we can investigate numerically. There are of course a large number of ways that we could construct such meaning spaces. We use a simple Markov chain model, in which meaning $1$ is inferred with probability $c$, and then meaning $m>1$ is inferred with probability $p_m$ if meaning $m-1$ is inferred, or with probability $q_m$ if meaning $m-1$ is not inferred.  Specifically, we choose
\begin{equation}
p_m = \frac{(1-\alpha) (m-1)^{\gamma} + \alpha c}{m^\gamma} \quad\mbox{and}\quad
q_m = \frac{\alpha c}{m^\gamma} 
\end{equation}
so that the probability that any particular meaning $m$ is inferred is $cm^{-\gamma}$, as previously. However, given that a meaning $m$ is inferred, there is an increased likelihood that meanings with a similar plausibility are inferred, depending on $\alpha$.  Specifically, if $\alpha=1$, $p_m=q_m$, which means that each meaning is inferred independently of any other meaning, which is the special case we treated before.  On the other hand, as $\alpha\to 0$, the probability $p_m \to 1$ as $m\to\infty$, which means that long strings of infrequent meanings appear together in this limit. 

\begin{figure}
\begin{center}
\includegraphics[width=\linewidth]{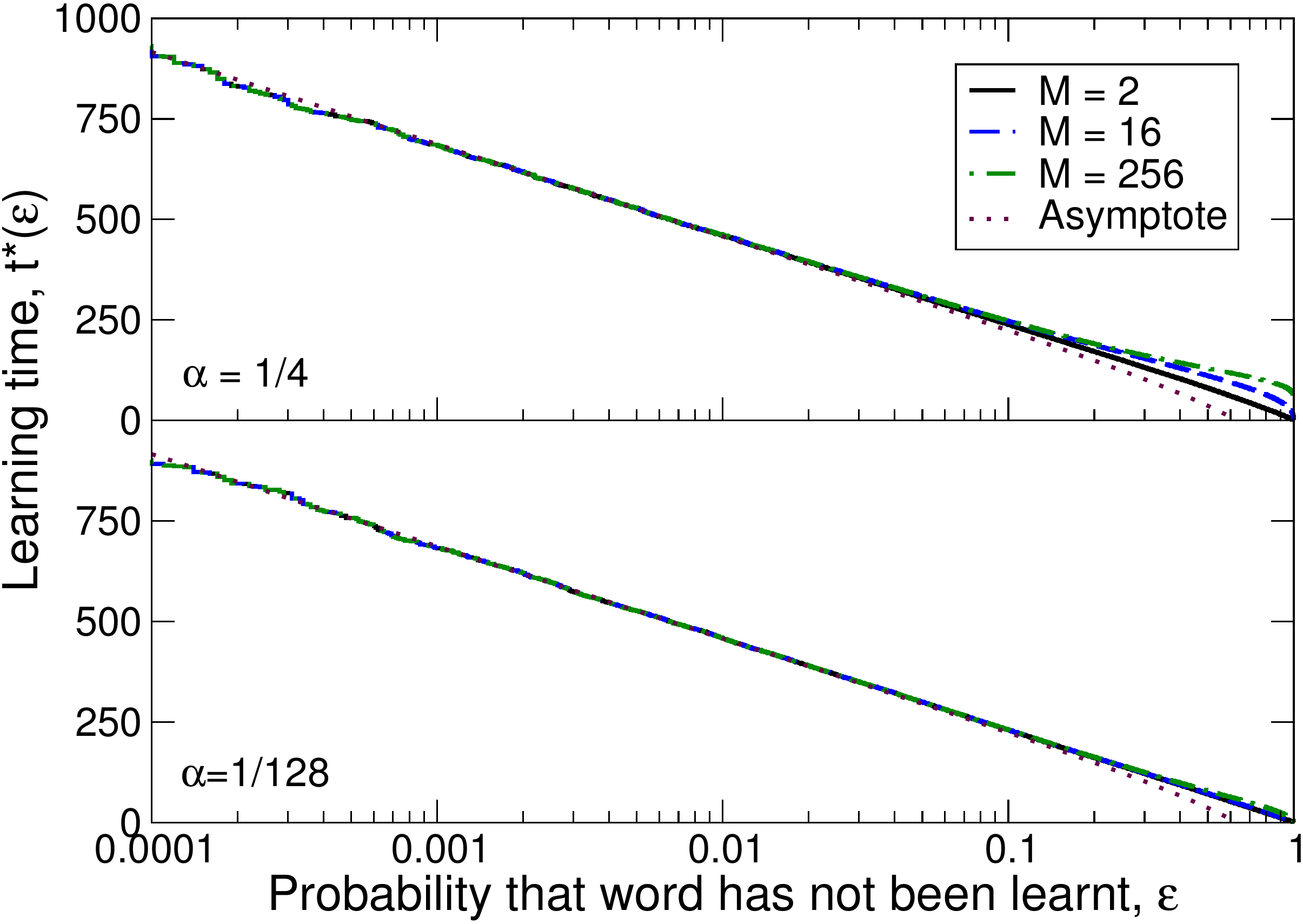}
\end{center}
\caption{\label{fig:oneword-power-domains} Time to learn a word with probability $1-\epsilon$ with confounder frequencies governed by the power-law with $c=0.99$ and $\gamma=0.01$. Meanings are correlated so that strings of neighbouring meanings are likely to appear, the degree of correlation being determined by $\alpha$. As in Fig.~\ref{fig:oneword-power}, learning time is independent of the number of incidental meanings, $M$. Furthermore, learning time is largely unaffected by correlations between meanings, as can be seen by comparing the upper and lower panels of this figure (within-episode correlations are substantially stronger in the lower panel), and indeed by comparing this figure to  Fig.~\ref{fig:oneword-power} (where there are no within-episode correlations): any differences from the uncorrelated case (Fig.~\ref{fig:oneword-power}) are confined to the large-$\epsilon$ regime.}
\end{figure}

As anticipated, we find that the learning time $t^\ast(\epsilon)$ is largely unaffected by the presence of correlations, even for small values of $\alpha$, as shown in Figure~\ref{fig:oneword-power-domains}. In particular, we find the same approximation that captured the behaviour of the statistically-independent case,  $t^\ast(\epsilon) \sim \ln(\epsilon)/\ln(c)$ as $\ln(\epsilon)\to 0$, also works well here: even when within-episode correlations are present, the slowest learners are those who retain the most plausible confounding meaning as an alternative hypothesis for an extended time.

\subsection{Correlated meaning inferences across episodes}
\label{sec:correlations_between_episodes}

Inferences mighty also be correlated in time: a meaning being inferred in one episode may enhance (or diminish) its probability of being inferred in the next episode, which would seem likely given that words are used in dialogues where certain themes and linguistic constructions recur \citep{pickering_04_towards}.  With such correlations, one can typically associate a timescale $\tau$ after which the correlation has decayed.  In this case, we would simply expect all learning timescales to be increased by a factor of order $\tau$.  This does not affect whether the learning time is infinite or not, unless $\tau$ itself is infinite.  

\section{Failure to infer the target meaning: relaxing Assumption 1}
\label{sec:target_failure}

In the foregoing models, we assumed that the word's true meaning, the target meaning, was inferred at every episode (our Assumption 1). While this significantly simplifies the formal analysis of cross-situational learning, it turns out \citep[contra e.g.][]{Gleitman90} not to be necessary for word learning to be possible. Rather, what is required is that the target meaning is the most likely to be inferred whenever the word is uttered: given enough exposures, this meaning will have been inferred more times than any arbitrarily high-probability incidental meaning, and a learner who identifies the meaning which is most strongly associated with the target word will learn the word \citep[as discussed in, e.g.,][]{Smith11}. Therefore, cross-situational word learning is possible as long as learners employ heuristics that reliably lead to the target meaning being the most plausible (over many episodes, not necessarily in any given episode).

In order to show that this is the case, we relax our Assumption 1: rather than always being inferred, we assume that the target meaning is inferred with probability $c^{\prime}$, and the probability of the incidental meanings decays according to a power law as before, with the probability of the $m$th incidental meaning being inferred on any episode being $a_m = c^{\prime} m^{-\gamma}$, $m$ ranging from 2 (for the most plausible incidental meaning) to infinity. $c^{\prime}=1$ corresponds to our standard model, where Assumption 1 is met and the target meaning is always inferred; $c^{\prime}<1$ corresponds to cases where Assumption 1 is violated and the target meaning is not always inferred, including cases (for $c^{\prime}<0.5$) where it is not inferred on the majority of exposures. Due to the power-law decay, regardless of the value of $c^{\prime}$, all incidental meanings are less likely to be inferred than the target meaning.

In order for the word's true meaning to be identifiable despite the possibility of the target meaning being absent on one or more exposures, we have to relax our model of cross-situational learning somewhat. Under these circumstances, a strictly eliminative cross-situational learner would be forced to conclude that the target word had {\it no} meaning, since for every meaning (including the target meaning) there will be episodes where that meaning is not inferred.  Instead, we assume a less strict form of cross-situational learning. Learners maintain a single candidate hypothesis about the word's meaning, which they revise after each exposure. If the candidate meaning is included in the set of meanings selected at exposure $i$, it is retained; otherwise, the single most frequent meaning included in the current exposure is selected as the new candidate (i.e. the meaning which has co-occurred most frequently, but not necessarily always, with the target word); in the event that there are multiple equally-frequent meanings, one is selected at random. In \citet{BSS10} we term a related form of cross-situational learning, where new candidate meanings are selected proportionally to their frequency of occurrence, {\it Approximate} cross-situational learning, to distinguish it from the maximally-strict eliminative form of cross-situational learning, and also from other forms of cross-situational which make less use of cross-situational statistics (e.g. the Minimal strategy from \citealt{BSS10}; Propose but Verify from \citealt{trueswell_13_propose}).

\begin{figure}
\begin{center}
\includegraphics[width=\linewidth]{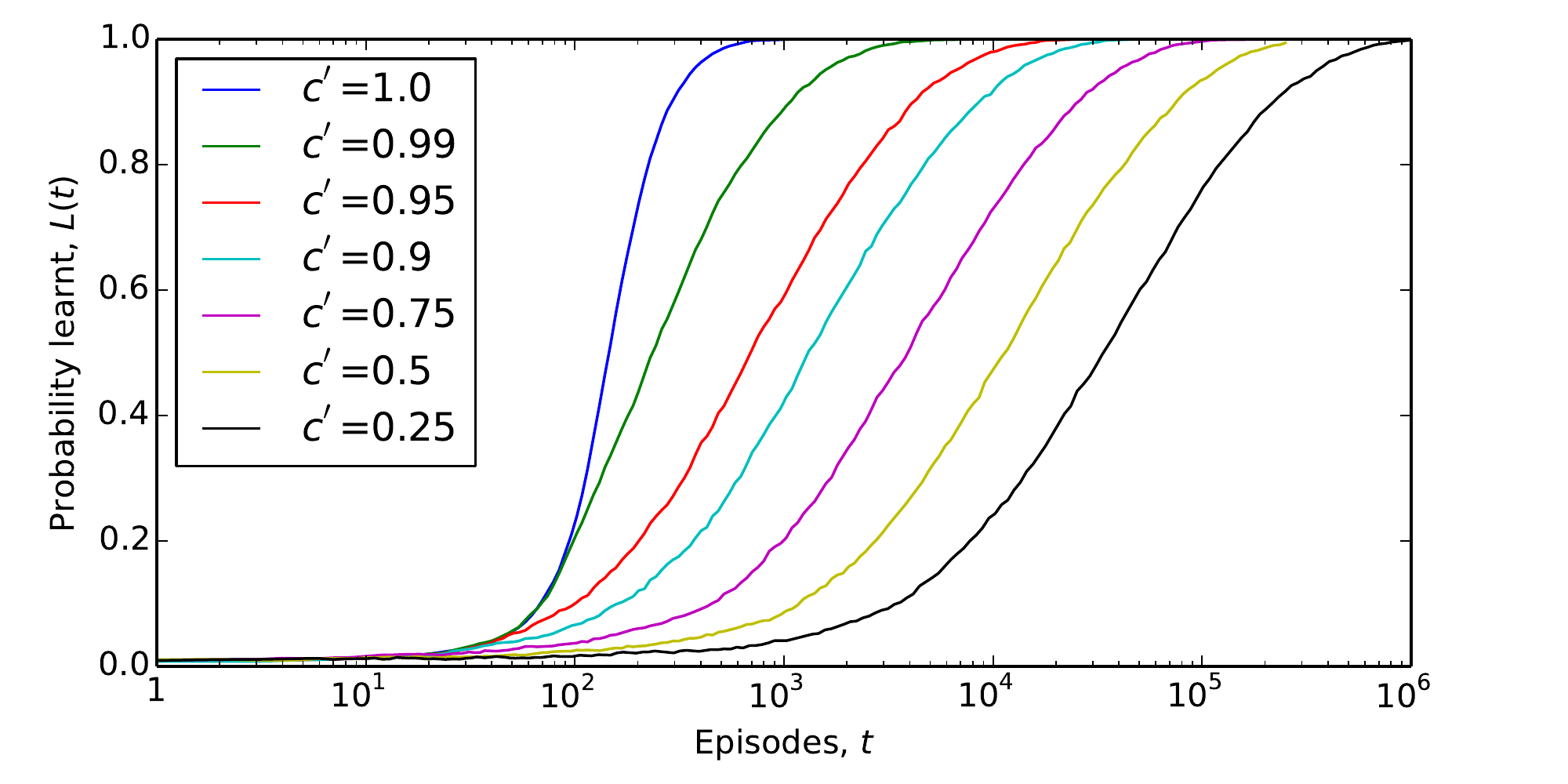}
\end{center}
\caption{\label{fig:missing_target} Proportion of 10,000 simulation runs in which the target word is learnt after a given number of exposures, with meaning frequencies governed by the power-law with various values of $c^{\prime}$, denoting the probability that the target meaning will be inferred on any given episode: $c^{\prime}=1$ corresponds to the case where the target is always inferred (satisfying our Assumption 1); learning is slower but still possible for $c^{\prime}<1$, corresponding to cases where the target meaning is not inferred on every trial, including for cases where $c^{\prime}<0.5$, i.e. where the target meaning is not inferred on the majority of trials. These results are for M=100; results for larger M are qualitatively similar.}
\end{figure}

As can be seen in Fig.~\ref{fig:missing_target}, the target word can still be learnt even if the target meaning is not reliably inferred (e.g. $c^{\prime}<1$) or even if it is not usually inferred ($c^{\prime}<0.5$) --- the less reliably the target is inferred, the slower learning becomes, but since the target meaning is still the single most probable meaning, eventually the cross-situational statistics will allow the learner to identify the correct meaning. In other words, cross-situational learning will be possible as long as the learner's heuristics ensure that the target meaning is the most likely to be inferred {\it in the long run}, even when the target meaning is never unambiguously provided and even when the target meaning is often not inferred.

\section{Discussion}

We make two central assumptions in our initial model: that the target meaning is {\it always} inferred (our Assumption 1), and that all incidental meanings have some non-zero probability of {\it not} being inferred, i.e. unlike the target meaning, they are not always inferred (our Assumption 2). As we show in Section \ref{sec:target_failure}, Assumption 1 is not required for cross-situational word learning to be possible: words can be learnt despite potentially infinite referential uncertainty even if a word's true meaning is not always, or not often, inferred. Our second assumption embodies Quine's observation, which we take to be necessary for {\it any} model of word learning, that incidental meanings which are in principle indistinguishable from the target meaning will block learning.  

We have focussed here on the learning of a single word, but the same techniques can be extended to the learning of large lexicons, using the techniques presented in our previous work \citep{BSS10,reisenauer_13_statistical}: in particular, \citet{reisenauer_13_statistical} consider the case where words are not learnt independently, such that learning one word can facilitate the learning of another, through the Mutual Exclusivity constraint.

Our idealised learner in our initial model employs the most powerful form of cross-situational learning possible, tracking the set of meanings which have consistently co-occurred with the target word. In Section \ref{sec:target_failure}, we consider a weaker form of cross-situational learning, where learners track not the ever-present set of word meanings, but simply maintain some estimate of the frequency of candidate word meanings. Still weaker versions of cross-situational learning are possible, and indeed seem to better characterise human cross-situational learning in some scenarios \citep{Medina11,trueswell_13_propose}: for instance, learners might simply retain a single preferred candidate meaning across exposures, persisting with or rejecting this hypothesis in the light of each exposure. We have previously shown, for the case of finite sets of incidental meanings, that weaker strategies tend to increase the time required to learn a lexicon, without introducing any qualitative shift in the conditions under which cross-situational learning is possible \citep{BSS10}. While exploring these mechanisms for the case of infinite meaning spaces is an area for future work, we expect that the picture presented here will in general hold: in particular, details of the exact cross-situational procedure applied are unlikely to influence whether learning time is finite or not.

Our treatment of meaning is rather minimal: meanings are simply discrete values drawn from a potentially infinite set. This extremely general treatment allows us to abstract away from any particular theory of word meaning, and as we mention when introducing the model, we expect that this same approach could be implemented in such a way that meanings could naturally be interpreted as being drawn from a hierarchically- and similarity-structured space, by manipulating the set of incidental meanings associated with the word's true meaning and the distribution from which those incidental meanings are drawn, such that incidental meanings which are related to the target are more likely to be inferred by the learner, and more general meanings are overall more likely to be inferred than more specific, restricted meanings. Note that, in the latter case, the cross-situational learner still requires it to be the case that the word's true meaning is the most likely to be inferred (see next paragraph) --- biases such as a size principle \citep{tenenbaum_99_bayesian} or Mutual Exclusivity, which we take to be part of the learner's heuristics, would be required to ensure that words with specific, restricted meanings could be successfully learnt. Verifying that this is the case would be a valuable addition to the initial model we outline here.

All of the models outlined above assume that a word's true meaning is the most likely meaning to be inferred on any one exposure (although it may not actually be inferred), and therefore, averaged over many exposures, will be the most frequently inferred meaning for that word, allowing cross-situational learning to take place. This assumption bears some discussion. Firstly, we have assumed in our models that the probability distribution over the target and incidental meanings remains constant over the course of learning, but this is not necessarily the case. For instance, for some word meanings which rely on learning {\it a priori} more probable meanings first and then applying Mutual Exclusivity, the target meaning may only be learnable once the learning of other words re-orders the plausibility of candidate meanings such that the target meaning becomes the most plausible remaining meaning. In other words, we expect that word learning will still be possible if the target word is {\it eventually} the most plausible. Secondly, the fact that words are socially learned and culturally transmitted provides some guarantee that the types of meanings learners tend to infer will indeed be the types of meanings that words come to have: a range of computational and experimental work shows that languages adapt to the biases of language learners as a result of their repeated learning and transmission \citep[e.g.][]{kirby_00_syntax,griffiths_07_language,kirby_08_cumulative}: while the structure of the environment also feeds in to this process \citep{perfors_14_language}, we should therefore expect that word meanings will become adapted to the heuristics that learners use to infer them; in the limiting case, a word whose meaning could not be inferred by learners would be guaranteed to change in meaning. While the process by which languages become adapted to the biases of language learners has been well studied in the general case, explicit models of the process for word learning would be worthwhile.

 Finally, to reiterate: although the results we present results require that, in the long run, the word's true meaning is the most likely to be inferred, there is still substantial work for cross-situational learning to do. Firstly, although the learner is more likely to infer the word's true meaning than any other, they still face substantial or indeed infinite referential uncertainty: on any given exposure, the target meaning is merely one of an infinite number of inferred meanings, and cross-situational learning is required to eliminate this uncertainty. Secondly, as we show in Section \ref{sec:target_failure}, the target meaning does not have to be reliably inferred to be learnt. Cross-situational learning therefore turns an apparently impossible learning problem --- inferring the meaning of a word from a series of exposures where each exposure offers infinitely many candidate meanings, possibly not including the true meaning --- into a tractable one.





\section{Conclusions}

We have established a very general formal foundation for the study of cross-situational learning, obtaining an exact expression for the probability, $L(t)$ that a word is learnt after $t$ exposures by an ideal cross-situational learner, valid for an arbitrary distribution of incidental meanings. We have also established criteria for when this learning time is finite, i.e. cross-situational learning is possible, based on the residual plausibility of an incidental meaning. When confounding meanings are inferred independently, the learning time is well estimated by the asymptotic formula $t^\ast = \ln \epsilon / \ln c$, where $c$ is the frequency of the most common incidental meaning. This approximation also works well when there are within-episode correlations between inferred meanings. Importantly for the debate in the cognitive sciences over word learning in the face of infinite uncertainty, we have identified at least one scenario in which referential uncertainty at every exposure is infinite, yet learning times are still finite. This finding suggests that the common intuition, that cross-situational learning is impossible in such circumstances, is incorrect.  Furthermore, we show that cross-situational learning is possible even if the learner's heuristics only impose very weak constraints on the plausibility ranking of possible meanings and even when the word's true meaning is not always (or even not often) inferred. This work therefore suggests that word learning heuristics can in principle be far weaker than previously suggested and still allow word learning, and therefore suggests that weaker, unreliable, probabilistic heuristics can play an important role; exploring such biases in real word learners is therefore a worthwhile empirical aim.

\bibliographystyle{elsarticle-harv}
\bibliography{xslrefs}


\newpage 
\appendix

\section{Derivation of the main result}

Here we show that Equations~(\ref{pmt}) and (\ref{iff}) and the statement that a word can be learnt in a finite time are equivalent. We recall that ${\cal K}(t)$ is the set of confounding meanings that have appeared in every episode until time $t$ and that ${\cal F}_M$ comprises the first $M$ meanings. We introduce
\begin{equation}
L_M(t) = \Pr\left[ {\cal F}_M \cap {\cal K}(t) = \emptyset \right] \;,
\end{equation}
the probability that the first $M$ meanings have all failed to appear at least once in the first $t$ episodes.  From the definition of conditional probability,
\begin{align}
L_M(t) &= \Pr\left[ M \not\in  {\cal K}(t) | {\cal F}_{M-1}  \cap {\cal K}(t) = \emptyset \right]\, \Pr\left[ {\cal F}_{M-1} \cap {\cal K}(t) = \emptyset  \right] \nonumber\\&= [ 1 - p_M(t) ] L_{M-1}(t)
\end{align}
where $p_M(t)$ is given by Eq.~(\ref{pmt}). This is valid for all $M>0$ if we take $L_0(t)=1$ for all $t$.  Hence,
\begin{equation}
L_M(t) = \prod_{m=1}^{M} [ 1 - p_m(t) ] \;.
\end{equation}
The probability that the target meaning has been disambiguated from all other meanings is
\begin{equation}
\label{Lt}
L(t) = \lim_{M\to\infty} L_M(t) \equiv \prod_{m=1}^{\infty} [ 1 - p_m(t) ] \;.
\end{equation}
Since $p_m(t)$ is a probability, and lies between zero and one, this infinite product exists (and furthermore itself lies between zero and one).

We are interested in the case where $L^\ast = \lim_{t\to\infty} L(t) = 1$, since this corresponds to the word having a finite learning time.  This fact follows straightforwardly from the definition of a limit, which is that for any $\epsilon$ sufficiently small, there exists a $t^*$ such that $L(t)>1-\epsilon$ for all $t>t^*$. The smallest $t^*$ for which this is true for any given $\epsilon$ corresponds to the learning time $t^*(\epsilon)$ defined in the main text.

To show that (\ref{iff}) is a necessary and sufficient condition for a finite learning time, our strategy is as follows. We shall first assume that (\ref{iff}) is true, and will then find that this implies $L^\ast=1$, and the learning time is finite. We then then consider the case where (\ref{iff}) does not hold, and find that $L^\ast<1$ as a consequence, and hence that the learning time is infinite.

In the first instance, where (\ref{iff}) is assumed, it follows that, for sufficiently large $t$, we have $\max_m\{p_m(t)\}$ arbitrarily small, and in particular less than unity.  We can then take the logarithm of (\ref{Lt}) and expand as a power series to obtain the convergent double series
\begin{equation}
\ln L(t) = - \sum_{m=1}^{\infty} \sum_{k=1}^{\infty} \frac{p_m(t)^k}{k} \;.
\end{equation}
Since all terms in this double series have the same sign, the order of the summation indices can be exchanged, and the resulting double series
\begin{equation}
\label{J}
\ln J(t) = - \sum_{k=1}^{\infty} \sum_{m=1}^{\infty} \frac{p_m(t)^k}{k}
\end{equation}
has the same limit: $L(t)=J(t)$ \citep{Scott66}. (By contrast, if either of these double series diverges, then so does the other).

Using the fact that
\begin{equation}
\sum_{m=1}^{\infty} p_m(t)^k \le \left( \sum_{m=1}^{\infty} p_m(t) \right)^k
\end{equation}
we find
\begin{equation}
\ln L(t) = \ln J(t) \ge - \sum_{k=1}^{\infty} \frac{1}{k} \left[ \sum_{m=1}^{\infty} p_m(t) \right]^k = \ln\left[ 1 - \sum_{m=1}^{\infty} p_m(t) \right]
\end{equation}
where the second equality follows because it is assumed at the outset that $\sum_{m=1}^{\infty} p_m(t)$ converges to a value smaller than unity.  Exponentiating both sides, we finally obtain the inequality
\begin{equation}
L(t) \ge  1 - \sum_{m=1}^{\infty} p_m(t)  \;.
\end{equation}
Since the condition (\ref{iff}) states that the series on the right-hand side of the previous expression vanishes as $t\to\infty$, we have that $L^* \ge 1$.  However, $L^*$ cannot exceed unity: hence $L^* = 1$ and the condition (\ref{iff}) is sufficient for a finite learning time.

We now show that (\ref{iff}) is also a necessary condition, i.e., that if (\ref{iff}) does not hold, the learning time is infinite.  We consider again (\ref{J}), and find that
\begin{equation}
\ln J(t) = - \sum_{k=1}^{\infty} \frac{1}{k} \sum_{m=1}^{\infty} p_m(t)^k \le - \sum_{m=1}^{\infty} p_m(t) \;.
\end{equation}
Now, if (\ref{iff}) does not hold, we must have the strict inequality
\begin{equation}
\lim_{t\to\infty} \ln J(t) < 0 \;.
\end{equation}
If the sum for $\ln J(t)$ converges for sufficiently large $t$, we have that $L(t)=J(t)$, and hence $L^\ast < 1$ when (\ref{iff}) does not hold. Hence, the learning time becomes infinite at some nonzero $\epsilon$ in this case.  On the other hand, if the sum for $\ln J(t)$ diverges at all finite times, so does the sum for $\ln L(t)$, which implies that $L^\ast=0$, which again indicates an infinite learning time.

\section{The case of logarithmic decay}

In the text we discuss the situation where inference frequencies decay logarithmically, e.g., $a_m = c \ln 2 / \ln(m+1)$, and mention that under this scenario the word is not necessarily learnt in finite time. By applying the integral test, one finds that the sum in (\ref{iff}) diverges at any time $t<\infty$, which in turn implies that the learning time diverges for any $\epsilon<1$.

\begin{figure}
\begin{center}
\includegraphics[width=\linewidth]{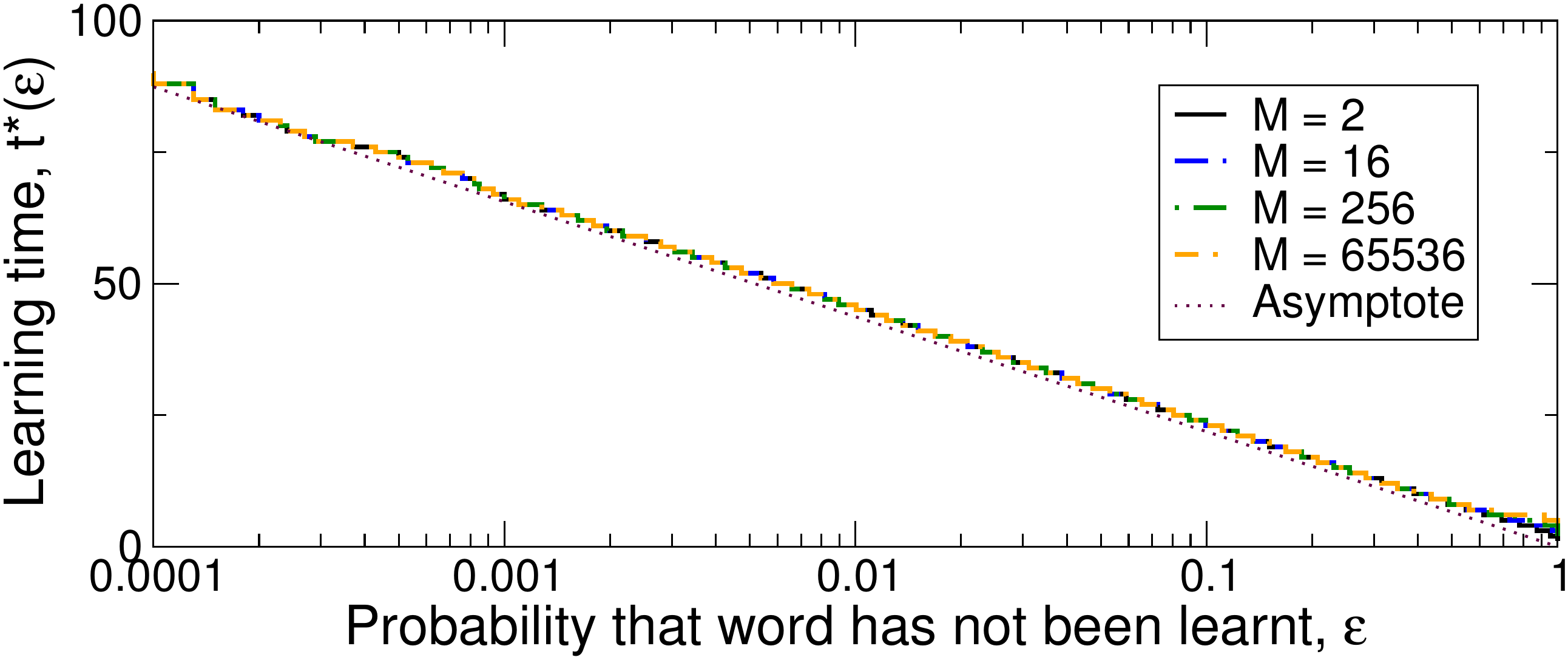}
\end{center}
\caption{\label{fig:oneword-log} Time to learn a word with probability $1-\epsilon$ with confounder frequencies governed by a logarithm $a_n = c \frac{\ln(2)}{\ln(n+1)}$, with $c=0.9$. Here the data for different sizes of the space of confounders, $M$, are indistinguishable except very close to $\epsilon=1$.  The learning times are close to the asymptotic result $t^\ast(\epsilon) \sim \ln(\epsilon)/\ln(c)$ even though formally in the limit $M\to\infty$ the learning time diverges for all $\epsilon$.}
\end{figure}

The way in which the learning time diverges in this example turns out to be quite subtle and somewhat revealing.  Suppose we keep only the first $M$ confounding meanings.  Then, from (\ref{Loft}) we have
\begin{equation}
L(t) \le \exp\left( - [c \ln 2]^t \sum_{m=1}^{M} \frac{1}{[\ln(m+1)]^t} \right)
\end{equation}
since $\ln(1-x)<-x$.  Replacing the sum with an integral, and making the change of variable $u = \ln m$ we find
\begin{equation}
L(t) \le
\exp\left( -[c \ln 2]^t \int_{\ln(2)}^{\ln(M+2)} \frac{{\rm e}^u}{u^t} {\rm d} u \right) \;.
\end{equation}
When $u$ is small, the exponential contribution to the integrand changes slowly, while the power-law part changes rapidly.  This means that the integral is weakly dependent on $M$ while it remains below some characteristic value.  In this regime, one can set ${\rm e}^{u} = 2$ (its value at the lower end of the integral), and   finds that the integral  {\it apparently} converges to 
\begin{equation}
\int_{\ln(2)}^{\ln(M+2)} \frac{2}{u^t} {\rm d} u = \frac{2}{t-1} \frac{1}{(\ln 2)^{t-1}} \left( 1 - \left[ \frac{\ln 2}{\ln (M+2)} \right]^{t-1} \right)
\end{equation}
which remains finite in the limit $M\to\infty$.  From this, one can estimate the behaviour of the learning time as $\epsilon\to 0$ as $t^\ast(\epsilon) = \ln(\epsilon)/\ln(c)$, which as previously is governed purely by the frequency of the most common confounder.  Numerical data (not shown) shown in Figure~\ref{fig:oneword-log}
correspond well with this estimate.

The integrand ${\rm e}^u / u^t$ has a minimum at $u=t$.  This provides an estimate of the value of $u$ at which we can no longer regard ${\rm e}^u$ as constant, and at which the integral begins to diverge.  The estimate of $t^\ast$ obtained above is therefore only valid if $u<t^\ast$ across the whole range of the integral, and in particular at the top end $u\approx \ln(M)$.  For this model we therefore find that the learning time will be apparently finite if $M < (\epsilon)^{1/\ln(c)}$. For the case $\epsilon=0.01$ and $c=0.9$, we would need to consider {\it at least} $10^{18}$ confounding meanings before probing the region where the integral diverges.

\end{document}